\documentclass[sigconf]{acmart}
\usepackage{marginnote}
\usepackage{booktabs}
\usepackage{marvosym}
\usepackage{enumitem}
\usepackage{caption}
\usepackage{float}
\usepackage{stfloats}
\usepackage{subcaption}
\usepackage[ruled,linesnumbered]{algorithm2e}
\usepackage{amsmath}
\usepackage{braket}
\usepackage{bm}
\usepackage[title]{appendix}
\usepackage{multirow}
\usepackage{color}
\AtBeginDocument{%
  \providecommand\BibTeX{{%
    \normalfont B\kern-0.5em{\scshape i\kern-0.25em b}\kern-0.8em\TeX}}}

\setcopyright{acmcopyright}
\copyrightyear{2023}
\acmYear{2023}
\acmDOI{10.1145/XXXXXXX.XXXXXX}

\acmConference[KDD'23]{Proceedings of the 29th ACM SIGKDD Conference on Knowledge Discovery and Data Mining}{August 6--10, 2023}{Long Beach, CA, USA.}
\acmBooktitle{Proceedings of the 29th ACM SIGKDD Conference on Knowledge Discovery and Data Mining (KDD '23), August 6--19, 2023, Long Beach, CA, USA}
\acmPrice{15.00}

\begin{document}

\title{Modeling Adaptive Fine-grained Task Relatedness  \\ for Joint CTR-CVR  Estimation}

\settopmatter{authorsperrow=4}

\author{Zihan Lin$^{\dagger}$}
\email{zhlin@ruc.edu.cn}
\affiliation{
  \institution{School of Information, Renmin University of China}
  \city{}
  \country{}
}
\author{Xuanhua Yang}
\email{xuanhua.yxh@alibaba-inc.com}
\affiliation{
  \institution{Alibaba Group}
  \city{Beijing}
  \country{China}
}

\author{Xiaoyu Peng}
\email{pengxiaoyu.pxy@alibaba-inc.com}
\affiliation{
  \institution{Alibaba Group}
  \city{Beijing}
  \country{China}
}
\author{Wayne Xin Zhao \textsuperscript{\Letter}}
\email{batmanfly@gmail.com}
\affiliation{
  \institution{Gaoling School of Artificial Intelligence, Renmin University of China}
  \city{}
  \country{}
}
\author{Shaoguo Liu\textsuperscript{\Letter}}
\email{shaoguo.lsg@alibaba-inc.com}
\affiliation{
  \institution{Alibaba Group}
  \city{Beijing}
  \country{China}
}
\author{Liang Wang}
\email{liangbo.wl@alibaba-inc.com}
\affiliation{
  \institution{Alibaba Group}
  \city{Beijing}
  \country{China}
}
\author{Bo Zheng}
\email{bozheng@alibaba-inc.com}
\affiliation{
  \institution{Alibaba Group}
  \city{Beijing}
  \country{China}
}
\thanks{$\dagger$ Work done during internship at Alibaba Group.}
\thanks{\textsuperscript{\Letter} Corresponding authors.}

\renewcommand{\authors}{Zihan Lin, Xuanhua Yang, Xiaoyu Peng, Wayne Xin Zhao, Shaoguo Liu, Liang Wang, Bo Zheng}
\renewcommand{\shortauthors}{Zihan Lin et al.}

\newcommand{\todo}[1]{\textcolor{blue}{#1}}
\newcommand{\ie}{\emph{i.e.,}\xspace}
\newcommand{\eg}{\emph{e.g.,}\xspace}
\newcommand{\aka}{\emph{a.k.a.,}\xspace}
\newcommand{\etal}{\emph{et al.}\xspace}
\newcommand{\paratitle}[1]{\vspace{1.5ex}\noindent\textbf{#1}}
\newcommand{\wrt}{w.r.t.\xspace}
\newcommand{\ignore}[1]{}
\newcommand{\our}{AdaFTR}
\begin{abstract}
In modern advertising and recommender systems, multi-task learning~(MTL) paradigm has been widely employed to jointly predict diverse user feedbacks (\eg click and purchase). 
While, existing MTL approaches are either rigid to adapt to different scenarios, or only capture coarse-grained task relatedness, thus making it difficult to effectively transfer knowledge across tasks. 

To address these issues, in this paper, we propose an \textbf{\underline{Ada}}ptive \textbf{\underline{F}}ine-grained \textbf{\underline{T}}ask \textbf{\underline{R}}elatedness modeling approach, \textbf{\our},  for joint CTR-CVR estimation.  Our approach is developed based on a parameter-sharing MTL architecture, and introduces a novel \emph{adaptive inter-task representation alignment} method based on contrastive learning.
Given an instance,   the inter-task representations of the same instance are considered  as positive, while the representations of another random instance are considered as negative. 
Furthermore, we explicitly model fine-grained task relatedness as the \emph{contrast strength} (\ie the temperature coefficient in InfoNCE loss) at the instance level. 
For this purpose, we build a relatedness prediction network, so  that it can predict the contrast strength for inter-task representations of an instance. In this way, we can adaptively set the temperature for contrastive learning in a fine-grained way (\ie instance level), so as to better capture task relatedness. 
Both \emph{offline evaluation} with public e-commerce datasets and \emph{online test} in a real advertising system at Alibaba have demonstrated the effectiveness of our approach. 

\end{abstract}



\maketitle

\section{Introduction}
Nowadays, recommender and advertising systems~\cite{yuan2014survey} have become increasingly important for discovering customer interest and providing personalized service in  online platforms. Basically, these systems conduct prediction tasks with logged user feedbacks~(\eg click and purchase) for inferring and tracking user's preference. Among the prediction tasks, estimating click-through rate~(CTR) and post-click conversion rate~(CVR)~\cite{ma2018entire,AutoHERI} have been widely studied in both research and industry communities.  These two  tasks are usually cast into supervised  classification~\cite{chen2016deep} or regression~\cite{effendi2017click}, requiring a capable  predictive model for accurate estimation. 

\begin{figure}[t]
\centering
\includegraphics[width=1\columnwidth]{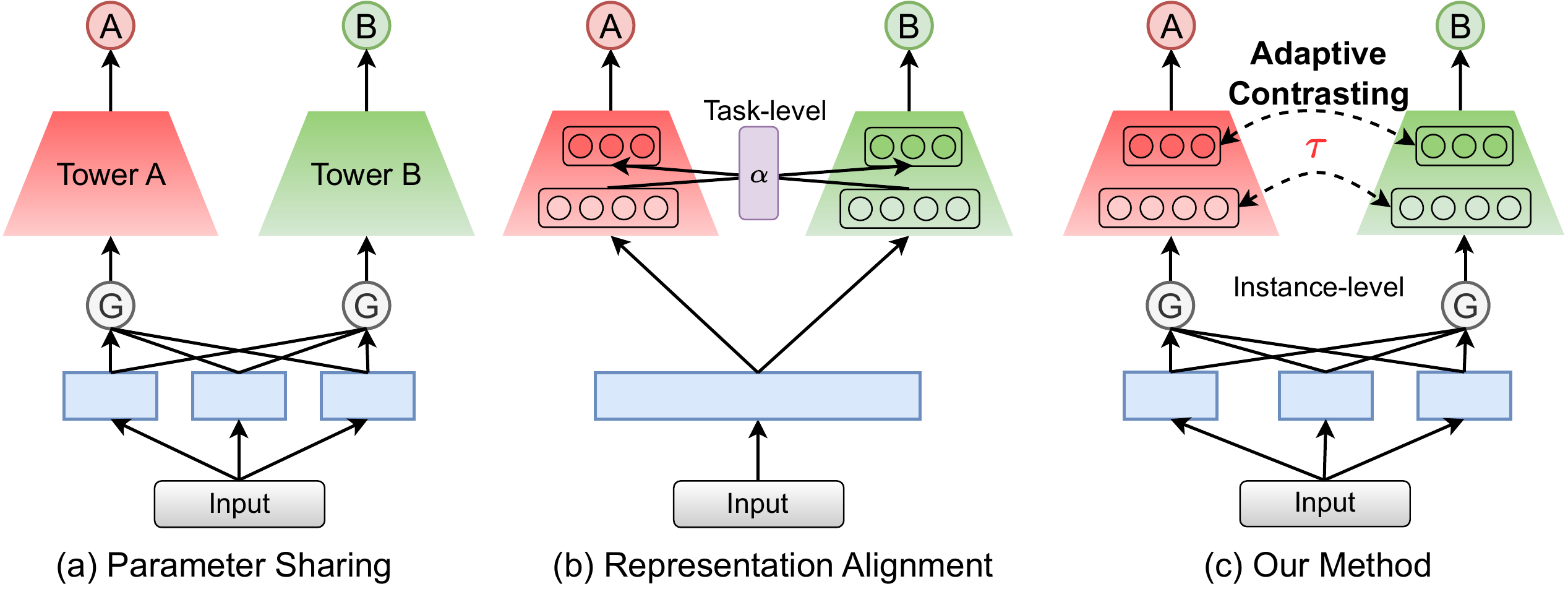} 
\caption{The comparison of different MTL architectures. Our approach actually combines the first two kinds of architectures, and further improves the task-specific networks with adaptive contrastive learning.}
\label{fig:intro}
\end{figure}

For the two kinds of behaviors~(\ie click and conversion) to be predicted, they are closely related to each other~\cite{ma2018entire}, and thus CTR and CVR can be jointly estimated by a unified model in a  multi-task learning~(MTL) paradigm~\cite{caruana1997multitask}. 
Recently, several studies have been developed to capture the inter-task relations for improving each individual task~\cite{AutoHERI,PLE,Cross-stitch}. 
Overall, existing work on joint CTR-CVR estimation can be divided  into two major categories. In the first category (\emph{parameter sharing}), the researchers seek to share model components across different tasks.  
As shown in Fig.~\ref{fig:intro}~(a),  they have mostly concentrated on designing or exploring the architecture for task-shared networks~\cite{Mmoe,PLE} (\eg layer sharing or expert sharing). Such architectures can directly transfer task knowledge via shared components, while they are less flexible to adapt to different tasks or datasets.
In the second category (\emph{representation alignment}), the studies aim to capture the associations between task-specific networks in either layer- or parameter-level~\cite{zou2022automatic,xiao2020lt4rec}. As shown in Fig.~\ref{fig:intro}~(b), it typically incorporates specific  components for capturing across-task associations~\cite{AutoHERI}, which require  additional maintenance costs  in online service.

In essence, the key to the joint CTR-CVR estimation  lies in the capturing \emph{task relatedness} so as to effectively transfer knowledge across tasks.
However, the task relation can be rather \emph{complicated} (\eg varying context, setting or distribution) in real application scenarios, and simply sharing layers or components (\eg  parameter sharing methods~\cite{Mmoe,PLE}) is too rigid to adapt to different scenarios. Furthermore, inter-task relatedness can be reflected very \emph{differently} for each individual instance, a coarse-grained relatedness model (\eg  representation alignment methods~\cite{AutoHERI}) is difficult to fully capture diverse correlating  patterns at the instance level. To summarize, existing approaches suffer from the two technical  issues, which cannot effectively solve the multi-task prediction problem.

\ignore{
For alleviating the above intractable issues, we propose to enhance each task with unsupervised representation learning paradigm to explicitly capture and utilize the relation between tasks. 
Contrastive learning recently has proved to have very promising results in unsupervised representation learning ~\cite{lin2022improving}.
which makes the representations uniformly distributed meanwhile aligning the representations of similar instances~\cite{wang2021understanding}. 
To this end, the positive and negative samples are constructed based on the underlying meaning. Motivated by these aspects, we purpose to capture the relationship between the task of CTR and CVR estimation with contrastive learning paradigm.}

To address these issues, in this paper, we propose an \textbf{\underline{Ada}}ptive \textbf{\underline{F}}ine-grained \textbf{\underline{T}}ask \textbf{\underline{R}}elatedness modeling approach, \textbf{\our},  for joint CTR-CVR estimation. To effectively transfer knowledge across tasks, we develop this approach  by combining both  parameter sharing and representation alignment.  Our backbone adopts a parameter-sharing MTL achitecture, where it places shared  layers  (embedding and experts) at the bottom and task-specific layers at the top. The major contribution lies in the proposed \emph{adaptive inter-task representation alignment} for capturing 
instance-level representation relatedness with contrastive learning. 
Compared with direct parameter sharing, contrastive learning can automatically learn the semantic alignment relation across different tasks according to the actual data distribution. In contrastive learning, 
it is important to select suitable positive and negative pairs for contrast.  Given an instance,  we  treat the inter-task representations of the same instance as positive, while the representations of another random instance as negative. 
 Unlike prior  contrastive learning methods~\cite{wang2022cl4ctr, lin2022improving}, 
we propose to explicitly model fine-grained task relatedness as the \emph{contrast strength} (\ie the temperature coefficient in InfoNCE loss~\cite{oord2018representation}) at the instance level. 
Specially, we build a representation relatedness prediction network, so  that it can predict the contrast strength for inter-task representations of an instance. In this way, we can adaptively set the temperature for contrastive learning in a fine-grained way (\ie instance level), so as to better capture inter-task relatedness. 

\ignore{
In this paper, we propose a novel \textbf{P}ersonalized \textbf{I}nter-task \textbf{CO}ntrastive Learning~(\textbf{\our}) framework which conducts contrastive learning between two tasks using in-batch negative samples and learnable instance-level temperatures. 
To solve the first challenge, we employ contrastive learning which regards the hidden representations of two tasks as positive pair to contrast. And the negative samples are introduced from the representations of other instance. 
To learn the relevance of tasks at instance level, we introduce an auxiliary network for two tasks to model the correlation of two labels on each instance. In this way, the learned relevance can be utilized to automatically control the instance-specific strength when contrasting the representation of two task, which is helpful to tackle the second challenge.
Creatively, we use the learned task relevance as the temperature parameter of InfoNCE~\cite{oord2018representation} loss to make the contrastive learning personalized for each data instance.
}

To verify the effectiveness of our  approach, we conduct both \emph{offline evaluation} with public e-commerce datasets and \emph{online test} in a real financial advertising system at Alibaba.  The major contributions of this paper can be summarized as follows:
\begin{itemize}[leftmargin=*]
    \item We present a hybrid architecture  that combine both parameter sharing and representation alignment for solving joint CTR-CVR  estimation, which can combine the mertis of the two kinds of MTL methods.  
    
    \item We propose an adaptive inter-task representation alignment method by explicitly modeling the inter-task representation relatedness at the instance level. It is model-agnostic in essence, and can be generalized to other MTL tasks. 
    \item Both offline and online experiments demonstrate that our approach significantly outperforms a number of competitive baseline models, and it indicates that our approach can effectively transfer knowledge across tasks by better modeling task relatedness. 
   \end{itemize}
\begin{figure*}[ht!]
\centering
\includegraphics[width=\textwidth]{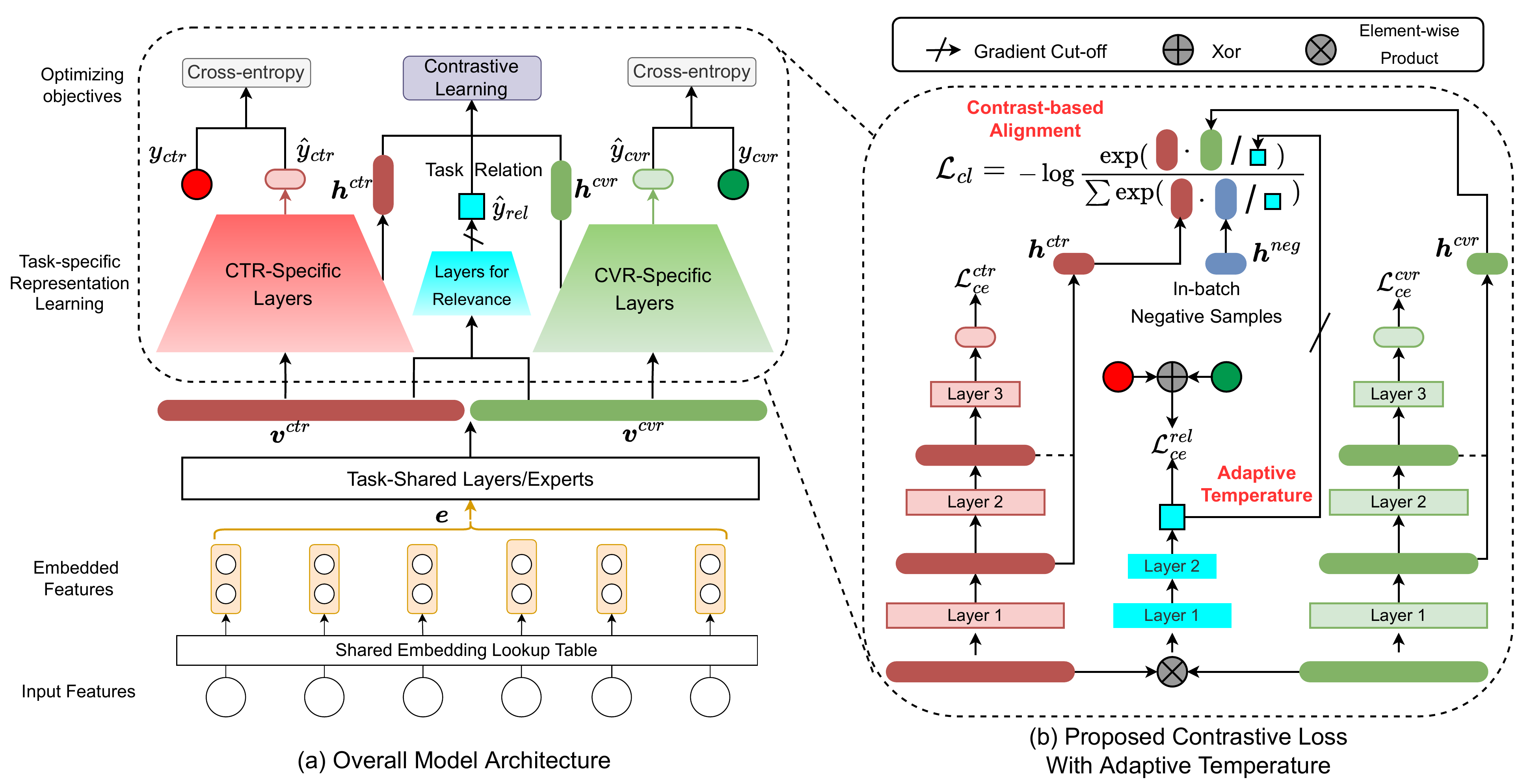} 
\caption{Overall framework of our proposed adaptive fine-grained relatedness method. The blocks in red, green, and light blue reflect the networks for CTR, CVR, and the relatedness, respectively. The white block represents the shared architecture. The rounded rectangles are identified as the representation vectors for contrastive learning.}
\label{fig:model}
\end{figure*}
\section{Methodology}

This section presents the proposed approach \our~  for joint CTR-CVR  estimation. The overall framework of \our~is shown in Figure~\ref{fig:model}. 
Our method is developed in a multi-task learning~(MTL) way, and introduces an adaptive inter-task  
contrastive learning method by modeling fine-grained task relatedness.   In what follows, we first describe the task formulation and the backbone architecture~(§~\ref{formulation-backbone}),  then introduce the proposed contrastive learning mechanism~(§~\ref{cl} -- ~\ref{rel}), and finally present the optimization strategy and complexity analysis~(§~\ref{opt}). 
 
\ignore{Without loss of generality, the method is applied on tasks of CTR and CVR estimation, while it can be easily extendable to other more tasks. 
 First, we describe the formulation and the backbone architecture~(§~\ref{formulation-backbone}). Then, we emphatically introduce the contrastive learning mechanism along with the learnable temperature parameter~(§~\ref{cl} -- ~\ref{rel}). Finally, we present the optimization strategy and complexity discussion~(§~\ref{opt}).
}

\subsection{Task Formulation and Model Backbone}\label{formulation-backbone}
In this work, we study the task of  joint CTR-CVR  estimation. 
As shown in prior studies~\cite{ma2018entire,AutoHERI}, CTR and CVR estimation are highly related when conducted on the same population of user traffic, and thus a MTL solution can potentially benefit  both CVR and CTR estimation. Note that, our approach \our~ is not   
limited to the two tasks,  and can be generally extended to other MTL problems.

\paratitle{Joint CTR-CVR  estimation}. CTR estimation  aims to predict whether  a user would click on a certain advertisement (or item) or not, and  CVR estimation aims to predict the probability of post-click conversion~(\eg view or purchase). To learn the prediction models, impression data will be logged by the application platform, represented as  a feature vector (consisting of user features, item features, and other context features) along with a user feedback label (whether or not a  user has performed actions such as clicking or purchasing). Formally, an impression record $r$ (\aka an instance) is denoted as $r=\langle \bm{x}, y_{ctr}, y_{cvr} \rangle$, where $\bm{x}$, ${y_{ctr} \in \{0,1\}}$ and ${y_{cvr} \in \{0,1\}}$ denote the feature vector, click label (CTR) and post-click conversion label (CVR), respectively. 
 
The joint  CTR-CVR estimation, aiming at learning a unified model that can predict both click and conversion labels, is denoted as: \begin{equation}\label{eq-general}
    \hat y_{ctr},\, \hat y_{cvr} = f(\bm{x} ; \Theta).
\end{equation}
where $\Theta$ represents the parameters that need to be learned.

\paratitle{Base architecture.} Our backbone adopts a typical MTL architecture, where it places  task-shared layers at the bottom and task-specific layers at the top. By introducing task-specific and shared layers, we can rewrite the general prediction formula in Equation~\eqref{eq-general}  as:
\begin{equation}
\begin{aligned} \label{eq:formal}
    \hat y_{ctr} &= f_{ctr}(g(\cdot)),\\
    \hat y_{cvr} &= f_{cvr}(g(\cdot)), 
\end{aligned}
\end{equation}
where $f_{ctr}(\cdot)$, $f_{cvr}(\cdot)$ and $g(\cdot)$ represents the CTR-specific layers, CVR-specific layers and task-shared layers respectively. As shown in Figure~\ref{fig:model}, the embedding features (denoted as $\bm{e}_r \in \mathbb{R}^{E}$), which take the one-hot feature vector ($\bm{x}_r$) for embedding table lookup, are the input of task-shared layers. 
To implement task-shared layers, we adopt the mixture-of-experts~(MoE) network~\cite{Mmoe,PLE}, where different tasks share the same set of expert networks $\{ g_i(\cdot) \}$. Further, it employs  
learnable weights to derive the output of the task-shared layers as:
\begin{equation}
\begin{aligned} \label{eq:MMoE}
 \boldsymbol{v}^{ctr} = \sum_{i=1}^{k}w^{(1)}(\bm{e}_r)_i \cdot g_i(\bm{e}_r), \\
 \boldsymbol{v}^{cvr} = \sum_{i=1}^{k}w^{(2)}(\bm{e}_r)_i \cdot g_i(\bm{e}_r),
\end{aligned}
\end{equation}
where $w^{(1)}(\cdot)$ and $w^{(2)}(\cdot)$ are the learnable combination weights based on the input for the CTR and CVR prediction tasks, respectively. Furthermore, $\boldsymbol{v}^{ctr}$ and $\boldsymbol{v}^{cvr}$ will be subsequently fed into the task-specific layers $f_{ctr}(\cdot)$, $f_{cvr}(\cdot)$ for generating the final predictions (Equation~\eqref{eq:formal}). Note that we do not use ``\emph{hard sharing}''~\cite{ruder2017overview} that directly shares network layers, but instead share expert components with learnable weights. It endows our approach with more flexibility in capturing complex relatedness patterns across tasks. 
 
\ignore{To our acknowledge, expert-sharing is a well-known mechanism to learn how the parameters are shared in those layers where several expert network~(MLP) are organized by the corresponding attention gates. In this way, the outputs of the task-shared layers( $\boldsymbol{v}^{ctr}$ and $\boldsymbol{v}^{cvr}$ ) of $g(\cdot)$ can be denoted as:
\begin{equation}
\begin{aligned}
 \boldsymbol{v}^{cvr} = \sum_{i=1}^{k}w_i(\bm{e}_r)g_i(\bm{e}_r), \\
 \boldsymbol{v}^{ctr} = \sum_{i=1}^{k}w_i(\bm{e}_r)g_i(\bm{e}_r),
\end{aligned}
\end{equation}
where $w_i(\cdot)$ is the learnable gates to control the routing on several experts. And $\boldsymbol{v}$ is the initial representation which will be sent to the task-specific network to further inject the unique knowledge of the corresponding task.
So far, many researches have devoted to designing the architecture of experts and gates to make them cooperate in phase~\cite{Mmoe,PLE}. }

\paratitle{Model learning}. After obtaining the predictions by Equation~\ref{eq:formal}, a straightforward optimization method~\cite{wang2022cl4ctr} is to learn  all the parameters with supervised binary cross-entropy loss:
\begin{equation}
\begin{aligned}
\mathcal L_{ctr} &= -y_{ctr} \cdot \log(\hat y_{ctr})-(1-y_{ctr}) \cdot \log(\hat y_{ctr}), \\
\mathcal L_{cvr} &= -y_{cvr} \cdot \log(\hat y_{cvr})-(1-y_{cvr}) \cdot \log(\hat y_{cvr}). \\
\end{aligned} \label{eq:bce}
\end{equation}
While,  inter-task relatedness is usually very complex to capture, and it is difficult to directly optimize the above loss in Equation~\eqref{eq:bce}, either sub-optimal or costly~\cite{Mmoe,PLE}. 
Thus, auxiliary loss~\cite{AutoHERI} has been proposed for enhancing the modeling of task relatedness for knowledge transfer. However, as discussed in Section~1, existing approaches are either less flexible to adapt to different scenarios~\cite{Mmoe}, or require more maintenance costs for additional componnts in online service~\cite{sheng2021one}.     
Based on the base architecture, we next present our solution by further modeling task relatedness in an adaptive, fine-grained way.  

\subsection{Inter-Task Representation Alignment}\label{cl}

In order to  enhance the task relatedness, we further propose to capture the semantic alignment relation across tasks. 
As aforementioned, task-specific layers, \ie $f_{ctr}(\cdot)$ and $f_{cvr}(\cdot)$, map shared data embeddings into task-specific representation space for prediction. 
Besides expert sharing, we further learn the alignment relation between task-specific representations for  knowledge transfer across tasks.  
Next, we introduce two  representation alignment methods, namely regularization and contrast based alignment.  

\paratitle{Regularization-based alignment.} As the first solution, we can directly 
align the hidden representations for CTR and CVR prediction tasks through the regularization based method. Given a record $r$, let $\boldsymbol{h}_r^{ctr}$ and $\boldsymbol{h}_r^{cvr}$ denote the vectors that are generated by hidden layers in $f_{ctr}(\cdot)$ and $f_{cvr}(\cdot)$, respectively, based on the feature vector for $r$. Then, we can set up a regularization loss to reduce the difference between $\boldsymbol{h}_r^{ctr}$ and $\boldsymbol{h}_r^{cvr}$: 
\begin{equation} \label{eq:reg}
    \mathcal{L}_{reg} = \ell (\boldsymbol{h}_r^{ctr}, \boldsymbol{h}_r^{cvr}),
\end{equation}
where $\ell(\cdot)$ can be set to various loss functions such as MSE or MAE.
Note that we place the regularization constraints on $\boldsymbol{h}_r^{ctr}$ and $\boldsymbol{h}_r^{cvr}$ instead of $\boldsymbol{v}_r^{ctr}$ and $\boldsymbol{v}_r^{cvr}$ (Equation~\eqref{eq:MMoE}), since $\boldsymbol{v}_r^{ctr}$ and $\boldsymbol{v}_r^{cvr}$ are already learned based on shared experts. 
Although regularization based alignment is widely used in existing multi-task learning work~\cite{lee2016asymmetric}, it is usually difficult to balance the task loss and  regularization loss:  a large weight on regularization loss will reduce the task-specific representation capacity, while a small weight cannot effectively fuse inter-task representations.  Such a loss balance becomes even more difficult when it involves a number of tasks. 


\paratitle{Contrast-based alignment.} To tackle the issue of regularization based alignment, we capture the task relatedness in representation space in a  \emph{inter-task contrastive learning} approach. Compared with regularization, contrastive learning is more \emph{capable} and \emph{flexible} in establishing the relatedness between different representation spaces~\cite{wang2020understanding}, which can better capture the semantic relatedness by leveraging rich instance associations. 
Specifically, given an impression record $r$, we treat its hidden representations from different tasks as positive pairs (\ie $\boldsymbol{h}_r^{ctr}$ and $\boldsymbol{h}_r^{cvr}$), while the hidden representations from another record $r'$ as negative samples. 
\ignore{\begin{equation}
    \mathcal{L}_{reg} = -\log(\boldsymbol{h}_x^{ctr}\cdot \boldsymbol{h}_x^{cvr}- \boldsymbol{h}_x^{ctr}\cdot \boldsymbol{h}_z^{cvr}).
    \label{eq: bpr}
\end{equation}
Here we use $\boldsymbol{h}_z^{cvr}$ as negative samples instead of $\boldsymbol{h}_z^{ctr}$ to ensure the same representation space. In this loss, the similarity between task representations is optimized and knowledge integration can be achieved between representations softly.
}
To implement such an alignment method, we adopt in-batch negatives for contrastive learning:  
\begin{equation}\label{eq:cl}
    \mathcal L_{cl} = -\sum_{r \in \mathcal R}\log\frac{\exp(\boldsymbol h_r^{ctr}\cdot \boldsymbol h_r^{cvr}/\tau)}{ \sum_{r'\in\mathcal B}\exp(\boldsymbol h_r^{ctr}\cdot \boldsymbol h_{r'}^{cvr}/\tau)},
\end{equation}
where $\mathcal B$ represents a batch of impression data and $\tau$ is the temperature parameter that controls the strength to contrast~\cite{wang2020understanding}. 
Note that here we take $\boldsymbol h_{r'}^{cvr}$ but not $\boldsymbol h_{r'}^{ctr}$ as the negative for $\boldsymbol h_{r}^{ctr}$, which is useful to enhance the fusion between inter-task representations. Following the classic BPR loss~\cite{rendle2009bpr}, we can also consider a simplified contrast variant where only one negative is involved:
\begin{equation}\label{eq:scl}
    \mathcal{L}_{scl} = -\log(\boldsymbol{h}_r^{ctr}\cdot \boldsymbol{h}_r^{cvr}- \boldsymbol{h}_r^{ctr}\cdot \boldsymbol{h}_{r'}^{cvr}).
\end{equation}

 In Equation~\eqref{eq:cl}, contrastive learning involves a temperature coefficient $\tau$,  which is usually set as a fixed value.  
It has been found that the temperature coefficient has a large effect on the performance of contrasting learning~\cite{zhang2021temperature,li2022curriculum}, and a fixed value can only lead to suboptimal performance. Considering this issue, we next seek  a more principled approach for setting the temperature coefficient. 
 

\subsection{Learning Adaptive Temperature by Modeling Representation Relatedness}\label{rel}


Instead of setting the temperature coefficient as a fixed hyper-parameter, we aim to develop an adaptive approach that automatically learns it for each instance (\ie an impression record).   Our solution is inspired by the finding that a smaller temperature usually indicates a more strict constraint (two representations are enforced to be more similar), and vice versa~\cite{wang2021understanding,zhang2021temperature}. 
Following this finding, it is intuitive that the temperature should be set to a smaller value when the inter-task representations of an impression record are more related.  
In order to better measure the inter-task representation relatedness for an instance, we propose to construct a relatedness prediction network and then employ the estimated relatedness to adaptively set the temperature coefficient.



\paratitle{Representation relatedness prediction}. Specifically, we additionally introduce an auxiliary network to learn the representation relatedness between the CTR and CVR tasks~(illustrated by light blue blocks in Figure~\ref{fig:model}). The relatedness prediction network takes the outputs of task-shared layers as inputs:
\begin{equation}
    \boldsymbol{v}^{rel} = \boldsymbol{v}^{ctr} \otimes \boldsymbol{v}^{cvr},
\end{equation}
where $\otimes$ represents element-wise product. 
Here, we use the representations $\boldsymbol{v}^{ctr}$ and $\boldsymbol{v}^{cvr}$ (Equation~\eqref{eq:MMoE}) learned based on shared expert networks. Intuitively, when two tasks are highly related, the derived mixture weights ($w^{(1)}(\cdot)$ and $w^{(2)}(\cdot)$ in Equation~\eqref{eq:MMoE}) should be also similar, thus leading to similar $\boldsymbol{v}^{ctr}$ and $\boldsymbol{v}^{cvr}$. 
Then, the representation for feature interaction denoted as $\boldsymbol{v}^{rel}$ will be subsequently transformed by MLP layers and a sigmoid function to generate the predicted relatedness $\hat y_{rel}$:
\begin{equation} \label{eq:rel_formal}
    \hat y_{rel} = \sigma\big(\text{MLP}(\boldsymbol{v}^{rel})\big).
\end{equation}
To train the relatedness prediction network, we employ the task  labels of CTR and CVR estimation (\ie $y_{ctr}$ and $y_{cvr}$) to derive the supervision signal as:  
\begin{equation} \label{eq:rel_construct}
    y_{rel} = \mathbb{I}[y_{ctr} = y_{cvr}],
\end{equation}
where $y_{rel}$ is set to 1 when $y_{ctr} = y_{cvr}$ and 0 otherwise. 
Intuitively, when the same impression record corresponds to the same label for both CTR and CVR tasks, their representations should be highly related. 
\ignore{The supervision signal of relevance network is determined by the labels of CTR and CVR tasks. Intuitively, The more instances that two tasks have the same labels~(both 1 or 0), the more relevant they are to each other. Therefore, We propose to generate the relevance label of each instance by: 
\begin{equation}
    y_{rel} = y_{ctr} \oplus y_{cvr},
\end{equation}
where $\oplus$ is xor operator\footnote{$\oplus$: $x \oplus y =\left\{
\begin{array}{cl}
1 &  x \ne y \\
0 &  x = y \\
\end{array} \right.$} 
}
To optimize the relatedness prediction network, we can also optimize the following  binary cross-entropy loss as:
\begin{equation}\label{eq:bce-rel}
    \mathcal{L}_{rel} = -y_{rel}\cdot \log(\hat y_{rel})-(1-y_{rel}) \cdot \log(\hat y_{rel}).
\end{equation}

\paratitle{Setting the temperature}.  After obtaining the predicted relatedness, we employ it to set the temperature parameter $\tau$ to control the strength in contrastive learning. 
 Here we consider using linear interpolation to control the temperature in an appropriate range:
\begin{equation}
    \tau_r = (\tau^U-\tau^L) \times (1-\hat y_{rel}) + \tau^L,
\label{eq:temperature}
\end{equation}
where $\tau^U$ and $\tau^L$ are the largest  and smallest value in a pre-set range  for the  temperature, respectively. In our experiments, we empirically find that  $\tau^U$ can be set as 1 and $\tau^L$ can be adjusted near to zero~(\eg 0.05). 
With the learned temperature, we update the inter-task contrastive learning as the follow equation:
\begin{equation}\label{eq-adaptive-cl}
    \mathcal L_{cl} = -\sum_{r \in \mathcal R}\log\frac{\exp(\boldsymbol h_r^{ctr}\cdot \boldsymbol h_r^{cvr}/\tau_r)}{ \sum_{r'\in\mathcal B}\exp(\boldsymbol h_r^{ctr}\cdot \boldsymbol h_{r'}^{cvr}/\tau_{r})},
\end{equation}
where $\tau_r$ is generated by Equation~\eqref{eq:temperature} for the impression record $r$. As we can see from Equation~\eqref{eq-adaptive-cl}, it can  adaptively tune the temperature according to the representation relatedness between tasks for different instances, thus leading to an improved capacity for representation alignment.

\subsection{Overall Optimization and Complexity}\label{opt}
In this part, we first discuss the model optimization and then analyze the model complexity. 
\subsubsection{Optimization}
To learn the whole framework, we consider a joint optimization objective that combines the aforementioned losses:   
\begin{equation}
    \mathcal L = \mathcal{L}_{ctr} + \mathcal{L}_{cvr} + \alpha \cdot \mathcal{L}_{rel} + \beta \cdot \mathcal{L}_{cl} + \lambda ||\Theta||_2,
\end{equation}
where $\mathcal{L}_{ctr}$ and $\mathcal{L}_{cvr}$ are the cross entropy loss defined in Equation~\eqref{eq:bce},   $\mathcal{L}_{rel}$ is the cross entropy loss for learning the adaptive temperature defined in Equation~\eqref{eq:bce-rel},  $\mathcal{L}_{cl}$ is the contrastive loss with the adaptive temperature defined in Equation~\eqref{eq-adaptive-cl},     $L2$ regularization is enforced for all the model parameters $\Theta$ to alleviate overfitting, and $\alpha$, 
$\beta$ and $\lambda$ are the combination weights for different losses.  
Note that we do not include the losses of $L_{reg}$ (Equation~\eqref{eq:reg}) and $L_{scl}$ (Equation~\eqref{eq:scl}) in the final objective, since they can be replaced with the more effective contrastive loss (\ie $\mathcal{L}_{cl}$), respectively. We will empirically discuss the effect of various optimization loss terms in Section~\ref{complexity}.  
During training, the gradients generated by the contrastive loss are only used to update the parameters for each task and the relatedness prediction network is optimized  by $\mathcal{L}_{cl}$.
The complete optimization process is shown in Algorithm~\ref{alg:training} where all the parameters are trained end-to-end. 
\begin{algorithm}[b]
    \caption{Training Algorithm for AdaFTR}
    \label{alg:training}
    \KwIn{Training dataset $\mathcal D$, range of temperature $\tau^{U}$ and $\tau^{L}$, learning rate $\gamma$, initial parameters for each task $\Theta$, initial parameters for task relatedness $\Omega$.}

    \While{Not converged}
    {
    Sample a batch of samples from $\mathcal D$\;
     \textbf{Get Model predictions}:\\
      Get $\hat y_{ctr}$ and $\hat y_{cvr}$ based on Eq.~\eqref{eq:formal}\;
      Get $\hat y_{rel}$ based on Eq.~\eqref{eq:rel_formal}\;
      Select hidden vectors $\boldsymbol h^{ctr}$ and $\boldsymbol h^{cvr}$ from outputs of task-specific layers\;
      \textbf{Relatedness parameter $\Omega $ optimization}:\\
      Construct $y_{rel}$ based on Eq.~\eqref{eq:rel_construct}\;
      Calculate $\mathcal{L}_{rel}$ based on Eq.~\eqref{eq:bce-rel}\;
      $\Omega \leftarrow \Omega-\gamma \nabla_{\Omega} \mathcal{L}_{rel}$ \;
      \textbf{Base parameter $\Theta $ optimization}:\\
      Calculate $\mathcal{L}_{ctr}$ and $\mathcal{L}_{cvr}$based on Eq.~\eqref{eq:bce}\;
      Calibrate the adaptive temperature $\tau_r$ based on Eq.~\eqref{eq:temperature}\;
      Sample negative instances $\boldsymbol h_{r'}^{cvr}$ from $\boldsymbol h^{cvr}$ \;
      Calculate $\mathcal{L}_{cl}$ based on Eq.~\eqref{eq-adaptive-cl}\;
      $\mathcal{L}_{model} \leftarrow weightSum(\mathcal{L}_{ctr}, \mathcal{L}_{cvr}, \mathcal{L}_{cl}, ||\Theta||_2) $ \;
      $\Theta \leftarrow \Theta-\gamma \nabla_{\Theta} \mathcal{L}_{model}$ \;
    }
\end{algorithm}

\ignore{\begin{equation}
    \mathcal L = \mathcal{L}_{ce}^{ctr} + \mathcal{L}_{ce}^{cvr} + \alpha \mathcal{L}_{ce}^{rel} + \beta \mathcal{L}_{cl} + \lambda ||\Theta||_2,
\end{equation}
where $\mathcal{L}_{ce}^{ctr}$ and $\mathcal{L}_{ce}^{cvr}$ are the cross entropy loss with the label of CTR and CVR respectively and $\mathcal{L}_{ce}^{rel}$ is the cross entropy loss to learn the personalized temperature. The contrastive loss is represented as $\mathcal{L}_{cl}$. We adopt L2 normalization on all the model parameters $\Theta$ to alleviate overfitting. To make all the objectives optimize synchronously, we introduce the hyper-parameters of $\alpha$, $\beta$ and $\lambda$ to rescale the proposed losses and the regularization term.
}

\subsubsection{Discussion on Complexity}\label{complexity}

\begin{table}[h]
\caption{The comparison on complexity between our proposed method and several representative MTL models. Here, $B$ is the training batch size  and $d$ is the dimension of hidden layers. For simplicity, we omit  other factors or ignorable costs. And we only consider the computation complexity of loss for training.} 
\label{tab:complexity}
\begin{tabular}{@{}ccccc@{}}
\toprule
Method      & MMoE   & Cross-Stitch   & AutoHERI  & AdaFTR~(Our)  \\ \midrule
Cat.     & \begin{tabular}[c]{@{}c@{}}Expert\\ Sharing\end{tabular} & \multicolumn{2}{c}{\begin{tabular}[c]{@{}c@{}}Representation\\ Alignment\end{tabular}} & Hybrid \\ \midrule
Train     & $\mathcal{O}(2B)$ & $\mathcal{O}(2B)$ &$\mathcal{O}(2B+L^2)$  & $\mathcal{O}(2B+B^2d)$\\
Deploy    & $\mathcal{O}(2Ld)$ & $\mathcal{O}(2Ld)$ & $\mathcal{O}(2Ld)$ & $\mathcal{O}(Ld)$        \\ \midrule
\#para. & $\mathcal{O}(2Ld)$ & $\mathcal{O}(2Ld+ 4L)$ &          $\mathcal{O}(2Ld + L^2)$ & $\mathcal{O}(2Ld)$ \\ \bottomrule
\end{tabular}
\end{table}

To compare our approach and other  representative methods, we list the time complexity for training/inference and the parameter number  in Table~\ref{tab:complexity}.  
For online serving, the efficiency of training and inference is an important factor to consider. 
Overall, our proposed approach follows a typical MTL paradigm,  thus having a similar time cost of training and inference with common methods~\cite{Mmoe,AITM}.  In our approach, we incorporate an additional auxiliary network to predict the representation relatedness, while it is only used during the training stage, which does not incur more inference cost.  In terms of training cost, most of the additional computation comes from the proposed inter-task contrastive learning; in terms of inference cost, our method can achieve a  higher efficiency for online services. Compared to other expert-sharing methods~\cite{PLE,Mmoe} or network fusion models~\cite{AutoHERI}, our framework is more lightweight, where the task-specific networks can be used independently for inference.      
Furthermore, the additional parameters are introduced by the relatedness prediction network, mainly consisting of MLP layers. As the additional network can be discarded after training, \our~ has comparable space cost with previous MTL models.

\ignore{
\our~only incorporates an additional auxiliary network to learn the task relevance, which is merely used at training stage. In this way, high efficiency is guaranteed without incurring any extra memory cost during the inference stage. In terms of time complexity, most of the additional computation comes from the proposed inter-task contrastive objective. We list the complexity on training loss and inference stage and the space complexity in Table~\ref{tab:complexity} where $B$ is the training batch size  and $d$ is the dimension of hidden layers. As we can find that our method can achieve high efficiency at inference with affordable additional time cost at training. Compared to other expert sharing methods~\cite{PLE,Mmoe} or network integration model~\cite{AutoHERI}, our framework is more lightweight where the redundant parameters can be removed when inference is only conducted on one task~(CTR or CVR)     
For the space complexity, the additional parameters are introduced by the relevance network which has several MLP layers with hundreds of neuron. As the additional parameters will not be saved after training, \our~ has comparable space cost with previous MTL models.
}
\section{Experiments}
To evaluate the effectiveness of the proposed \our, we conduct extensive experiments on both public dataset and industrial dataset. In-depth analysis is presented to give further understanding. Then, we further deploy it on real advertising system to conduct A/B test. 
\subsection{Experimental Setups}
\subsubsection{Datasets} 
We conduct the offline experiments on two industrial datasets. The first one is \textbf{Ali-CCP}~\cite{ma2018entire} which is a public dataset with 23 categorical features. We follow the original training/test splitting. The second dataset, \textbf{Ecomm-Ads}, is collected from a large-scale adverting system for online shopping. We collect the users' feedbacks for 8 days and the model is trained on the data of first seven days and tested on the data of last day. The dataset contains over one hundred categorical features and three behaviors~(\eg click, view and conversion). The statistics of two experimental datasets are listed in Table~\ref{tab:datasets}.

\begin{table}[]
\caption{Statistics of two experimental datasets}
\begin{tabular}{@{}ccccc@{}}
\toprule
Dataset   & \multicolumn{1}{c}{\#impression} & \multicolumn{1}{c}{\#click} & \#view                        & \multicolumn{1}{c}{\#conversion} \\ \midrule
Ali-CCP   & 85 M                       & 3.3 M                   & \multicolumn{1}{c}{-}         & 18 K                           \\
Ecomm-Ads & 0.7 B                      & 20 M                  & \multicolumn{1}{r}{8.7 M} & 1.1 M                        \\ \bottomrule
\end{tabular}
\label{tab:datasets}
\end{table}

\subsubsection{Evaluation Metrics}
We adopt two widely used metrics, Area Under ROC Curve~(\textbf{AUC}) and group AUC~(\textbf{GAUC})~\cite{he2016ups}, for all the offline experiments.  AUC indicates the probability of ranking positive samples higher than negative samples. GAUC is the extended metric of AUC for recommendation or advertising where the samples are grouped based on user and the final score is the weighted average of the AUC scores on each user. It is defined as:
\begin{equation}
    \text{GAUC} = \frac {\sum_{u} w_{u} * AUC_{u}}{\sum w_{u}},
\end{equation}
where $AUC_{u}$is the AUC score over the samples of user $u$ and $w_u$ is the weight for user $u$. We set all the weights to 1 in our experiments\footnote{Due to limited sapce, GAUC is only reported on Ecomm-Ads dataset as it performs much similar as AUC on Ali-CCP dataset.}.

\subsubsection{Compared Methods}
We compare the proposed \our~with several state-of-the-art multi-task learning methods: 
\begin{itemize}[leftmargin=*]
    \item \textbf{Single-DNN}~\cite{covington2016deep} trains independent MLP networks for different tasks.
    \item \textbf{Shared-Bottom}~\cite{caruana1997multitask} makes the embedding table and bottom MLP layers totally shared for both tasks and the upper layers are separately trained with each label.
    \item \textbf{Cross-stitch}~\cite{Cross-stitch} introduces learnable weights between different networks to fuse the knowledge of each task.
    \item \textbf{MMoE}~\cite{Mmoe} utilizes several shared experts as the bottom layers and learns an attention gate for each task to control the activation on all model parameters.
    \item \textbf{PLE}~\cite{PLE} additionally divides the experts into both shared ones and independent ones to better retain the individuality of tasks.
    \item \textbf{AutoHERI}~\cite{AutoHERI} introduces automated hierarchical representation integration between the individual networks and iteratively learns the connection weights on each layer.
\end{itemize} 

\subsubsection{Experimental Details}
To ensure fair comparison, we fix the feature embedding size to 8 and $\lambda$ to 1 for all the methods. The network for each task is a three-layered MLP and we set the hidden sizes as [128, 64, 32] on Ecomm-Ads dataset and [64, 32, 16] on Ali-CCP dataset due to their different data scales. The number of experts is set to three for all the expert-sharing methods and the first MLP layer is shared for other models. We use Adam optimizer with the batch size of 1024 and set the learning rate to 0.0005 consistently. Other hyper-parameters of baselines are carefully searched and tuned to the best value. The hyper-parameter $\alpha$ and $\beta$ are set to 1 and 0.01 based on grid search. The lower bound of $\tau$ is set to 0.05. The best value of hyper-parameters is consistent on both datasets according to our experiments. And we select the output of first task-specific layer as $\boldsymbol{h}_r$ in Equation~\eqref{eq-adaptive-cl} to contrast. All the models are trained for three times repeatedly and the average results over three runs are reported. The Ecomm-Ads dataset and source code will be public after the paper is accepted. For the convenience of reproducibility, we list all the hyper-parameters of our method in Table~\ref{tab:notation}.
\begin{table}[]
\caption{The setting of hyper-parameters in our experiments on both datasets and corresponding explanation  for reproducibility.}
\label{tab:notation}
\begin{tabular}{@{}c|l@{}}
\toprule
Hyper-parameters & Explanation \\ \midrule
$\tau^{U}=1$ & The upper bound of learned $\tau$.         \\ \midrule
$\tau^{L}=0.05$  & The lower bound of learned $\tau$.            \\ \midrule
 $\alpha=1$ & The coefficients for $\mathcal{L}_{rel}$.              \\ \midrule
 $\beta=0.01$ & The coefficients for $\mathcal{L}_{cl}$.             \\ \midrule
$\lambda=1$  & The coefficients for L2 Norm.             \\ \midrule
 $B=1024$ & The batch size for training.             \\ \midrule
$\gamma=0.0005$ & The learning rate for training.             \\ \midrule
 $L=3$ & The layers of task-specific network.            \\ \bottomrule
 $E=8$  & The dimension of feature embedding.           \\ \bottomrule
$d=128$  & The dimension of hidden layers.            \\ \bottomrule
\end{tabular}
\end{table}

\begin{table*}[ht!]
\caption{Experimental results on Ali-CCP and Ecomm-Ads datasets where the best one is \textbf{bolded} and the runner-up is \underline{underlined}. Symbol $^*$ indicates that the improvement is statistically significant where the $p$-value is smaller than 0.05. For the CVR task, the improvement over 0.1 on AUC is considered to be pratically valuable~\cite{ma2018entire} and \our~performs comparably on the CTR task.}
\begin{tabular}{@{}c|cc|cccc@{}}
\toprule
Dataset       & \multicolumn{2}{c|}{Ali-CCP}    & \multicolumn{4}{c}{Ecomm-Ads}                                                                                  \\ \midrule
Task          & {Click}       & {Conversion}                           & \multicolumn{2}{c}{Click}                            & \multicolumn{2}{c}{Conversion}                          \\
Metric        & AUC                      & AUC             & AUC                             & GAUC                       & AUC                        & GAUC                       \\ \midrule 
Single-DNN    & \textbf{61.977}                             & 59.525                                   & \textbf{83.673}        & \textbf{78.286}           & 86.929                    & 82.339                    \\
Shared-Bottom & {\underline{61.975~(-0.002)}}                   & 61.643~(+2.118)                           & 83.622~(-0.051)       & 78.219~(-0.067)          & 87.354~(+0.425)          & {\underline{ 83.027~(+0.688)}}    \\
Cross-Stitch  & 61.936~(-0.039)                         & 61.782~(+2.257)                       & 83.641~(-0.032)       & 78.235~(-0.051)          & 87.376~(+0.447)          & 83.008~(+0.669)          \\
MMoE          & 61.843~(-0.132)                 & 60.111~(+0.586)                  & 83.642~(-0.031)       & 78.256~(-0.030)          & {\underline{87.393~(+0.464)}}    & 83.006~(+0.667)          \\
PLE           & 61.889~(-0.086)                           & 60.499~(+0.974)                      & 83.660~(-0.013)       & 78.265~(-0.021)          & 87.188~(+0.259)          & 82.798~(+0.459)          \\
AutoHERI      & 61.968~(-0.007)                         & 60.387~(+0.862)                             & 83.657~(-0.016)       & {\underline{78.277~(-0.009)}}    & 87.335~(+0.406)          & 82.971~(+0.632)          \\
\our         & 61.886~(-0.089)                   & \textbf{61.928~(+2.403)$^*$}           & {\underline{ 83.667~(-0.006)}} & \textbf{78.286~(+0.000)} & \textbf{87.569~(+0.640)}$^*$ & \textbf{83.285~(+0.946)$^*$} \\ \bottomrule
\end{tabular}

\label{tab:overall-exp}
\end{table*}
\subsection{Overall Performance}
The performance comparison of the proposed \our~and other baseline methods on two datasets is shown in Table~\ref{tab:overall-exp}. We find several insightful observations:

(1) Compared to Single-DNN method which trains independent networks for each task, the other multi-task learning methods gain some improvements on CVR task. While the performance of CTR task decreases slightly on both datasets. This situation is mainly due to the imbalance between positive and negative label~\cite{ma2018entire} in real-world advertising scenarios. The positive labels for CVR task are extremely sparse compared to CTR task~(\eg 183:1 on Ali-CCP dataset and 18:1 on Ecomm-Ads dataset) which makes the learning of CTR more stable and hard to be improved. Therefore, the networks for CTR task can be learned adequately and the introduction of CVR task is meaningless. 
Generally speaking, the task with sparser label (\eg CVR) can obtain more benefits from the additional knowledge that introduced by other tasks. 
In this way, the effectiveness of multi-task learning is validated as CVR is usually the more important objective in most of industrial scenarios~\cite{AutoHERI}.

(2) Among the multi-task learning baselines, we find that MMoE performs best on Ecomm-Ads dataset which shows the effectiveness of expert-sharing and gating mechanism in controlling the feature integration for multiple tasks. However, Shared-Bottom and Cross-stitch perform better on Ali-CCP dataset. A possible reason is that those expert-sharing methods have complex parameters to learn for the attention gates, which can not be well optimized with limited data. Especially in PLE, the experts and gates for only CVR task can not be well-trained as the distribution of label is skew and supervised signals for CVR task are deficient.

(3) Finally, as for \our, we can observe comparable results on CTR task and significant improvements on CVR task compared to baseline methods~(increasing over 0.1 is significant for advertising). The superiority of \our~is consistent over two datasets. The advantage is mainly brought by the inter-task contrastive learning paradigm, which explicitly captures the potential shared knowledge across tasks to help the representation learning for the task of CVR. Furthermore, the adaptive temperature parameters make the contrastive representation integration at instance-level, which can better capture task relatedness compared with naive network connection. In addition, our method gains more improvement on sparser dataset~(\eg Ali-CCP) which shows the great potential of contrastive learning in multi-task scenarios.

\subsection{Further Study}
To give in-depth understanding of the proposed \our, we conduct a series of detailed experiments to analyse the effectiveness. We only report the results on Ecomm-Ads dataset as it is large-scale and industrial.

\subsubsection{Ablation Study.} 
\begin{table}[]
\caption{Performance comparison of different variants. Each component is indispensable in \our.}
\begin{tabular}{ccccc}
\toprule
\multicolumn{1}{c}{\multirow{2}{*}{Model}} & \multicolumn{2}{c}{Click} & \multicolumn{2}{c}{Conversion} \\
\multicolumn{1}{c}{}                       & AUC         & GAUC        & AUC            & GAUC          \\ \midrule
MMoE                                       & 83.642      & 78.256      & 87.393         & 83.008        \\ \midrule
\our                                      & 83.667      & \textbf{78.286 }     & \textbf{87.569}         & \textbf{83.285}        \\
\enspace w/o adaptive $\tau$                       & 83.659      & 78.266      & 87.493         & 83.283        \\
\enspace w/o learnable $\tau$                          & 83.637      & 78.221      & 87.381         & 83.083        \\
\enspace w/o negative    &  83.628           & 78.175            & 87.344               & 82.987              \\
\enspace w/o contrastive                            & \textbf{83.670}      & 78.249      & 87.326         & 83.034        \\ \bottomrule

\end{tabular}

\label{tab:exp-ablation}
\end{table}
There are several components in the proposed \our. To study the effectiveness of each parts, we compare our method with four variants: 
\begin{itemize}[leftmargin=*]
\item \emph{w/o adaptive $\tau$} removes adaptive temperature parameter and utilize one learnable parameter as the $\tau$ in Equation~\eqref{eq-adaptive-cl} for all the instances. 
\item \emph{w/o learnable $\tau$} removes learnable temperature parameter and set $\tau$ in Equation~\eqref{eq-adaptive-cl} to 0.05 based on statistical relevance. 
\item \emph{w/o negative} removes the negative samples in contrastive learning and replace the contrstive loss to $\mathcal{L}_{scl}$ in Equation ~\eqref{eq:scl}. 
\item \emph{w/o contrastive} removes the contrastive learning completely and aligns the representations with Equation ~\eqref{eq:reg}. 
\end{itemize}
Their results are reported in Table \ref{tab:exp-ablation}. Through the comparison, we can find that learning adaptive temperature is a useful strategy for inter-task contrastive learning, which captures the relatedness of tasks on each instance to make fine-grained representation integration. Meanwhile, it can be observed that learned temperature has better performance than manually selected parameter. It reveals the great potential of automatic architecture search in the field of recommendation and advertising. Besides, it has shown that simply optimizing the distance between task representations without negative samples yields a large performance drop, which further validates the effectiveness of contrastive learning paradigm.

\subsubsection{Using Different Hidden Vector $\boldsymbol h$ for Contrasting.} 
\begin{figure}[]
    \begin{minipage}[t]{0.75\linewidth}
		\centering
		\includegraphics[width=1\textwidth]{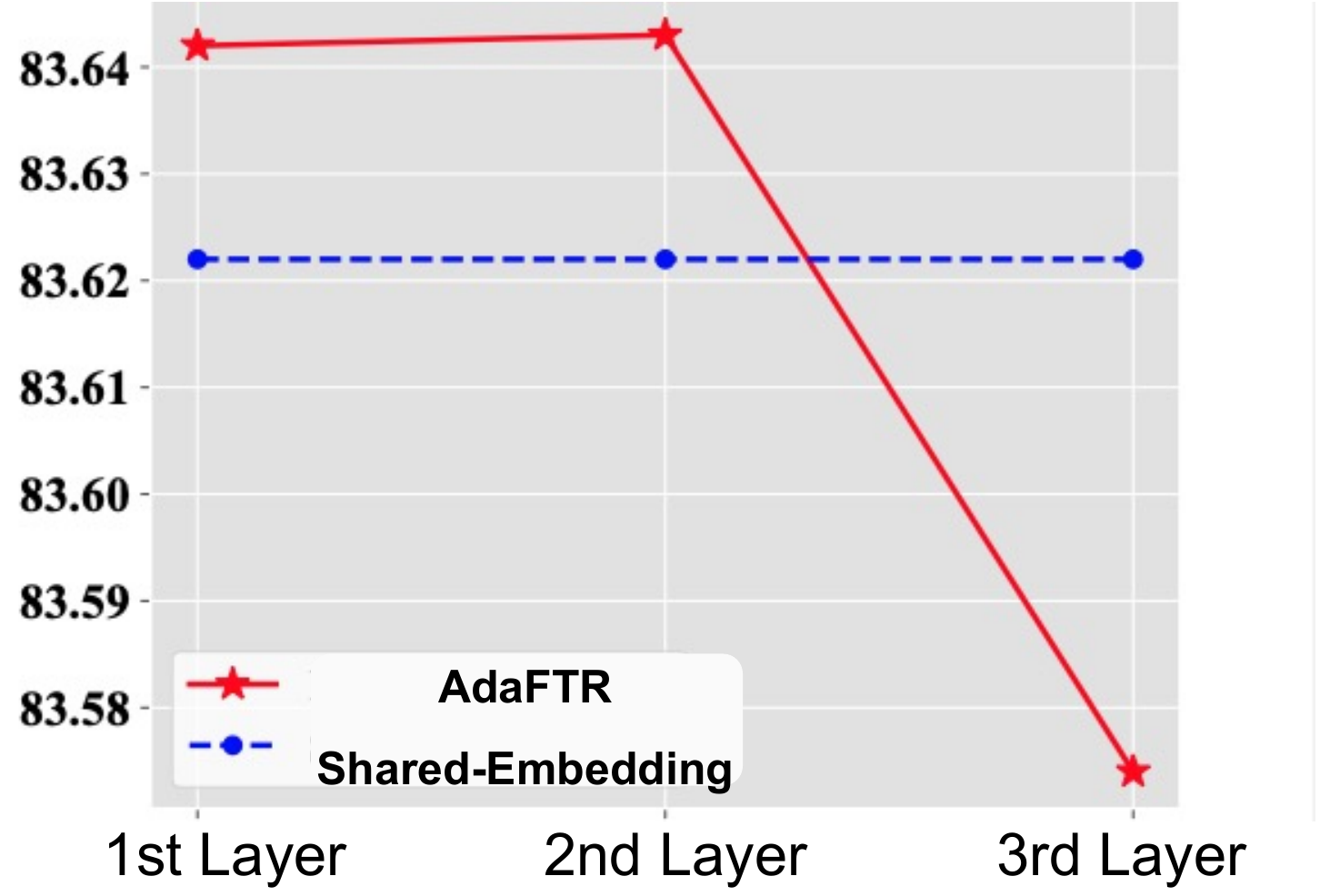}
           \subcaption{AUC results on the task of CTR}
		\label{fig:exp-hiddenlayer-clk}
	\end{minipage}
	\begin{minipage}[t]{0.75\linewidth}
		\centering
		\includegraphics[width=1\textwidth]{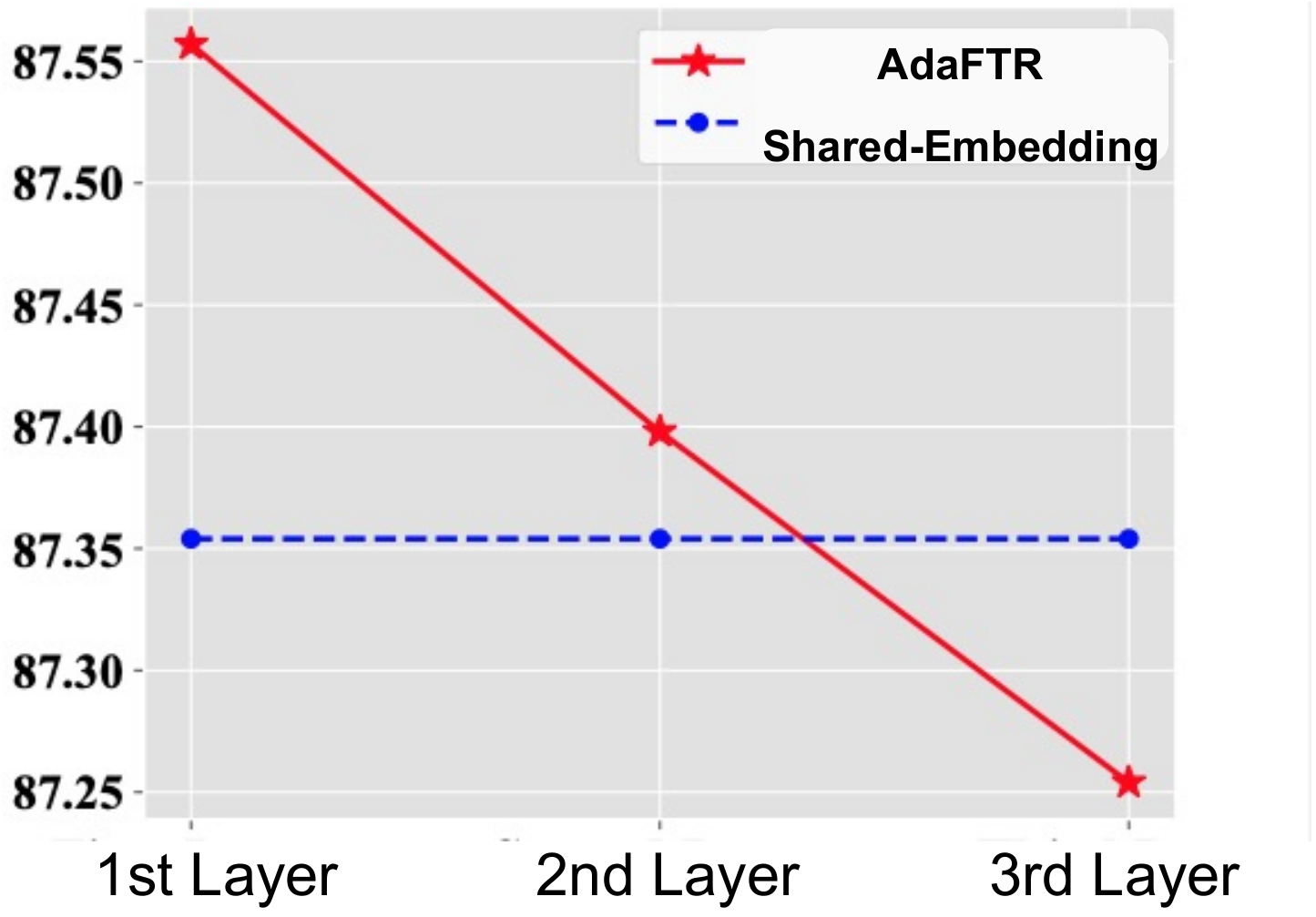}
           \subcaption{AUC results on the task of CVR}
		\label{fig:exp-hiddenlayer-pay}
	\end{minipage}
    \caption{Performance comparison w.r.t selecting hidden vectors from different layers to contrast. The results on both tasks are illustrated at left and right, respectively.}
    \label{fig:exp-hiddenlayer}
\end{figure}

In the proposed method, the task representations are generated by the individual MLP networks. To study how \our~performs when contrastive learning is conducted on different choices of task representations, we select the output of the first, second and third MLP layer in the task-specific network as $\boldsymbol h$ in Equation~\eqref{eq-adaptive-cl} to conduct the proposed adaptive contrastive learning. The results are shown in Figure~\ref{fig:exp-hiddenlayer}, which show that the performance on CVR task drops gradually when we select the outputs of higher layer to contrast. Similarly, the performance on CTR task is damaged severely when we contrast the outputs of the third layer. It is suggested that we should preserve some task-independent parameters to capture the specification of different tasks and achieve knowledge integration between the bottom layers for more benefits.

\subsubsection{Applying \our~with Different Backbones.}
As our method is structure-agnostic, we further apply it with different backbones to test the robustness of our method. Three classic MTL structures are selected which have shown their effectiveness in multiple behaviors modeling. The results are reported in Table \ref{tab:exp-general}. As we can observe, the proposed method can outperform the base model consistently on CVR task which further indicates the feasibility and applicability of our framework. Moreover, compared to MMoE, the relative improvement on PLE is more significant on CVR task. The possible reason is that PLE has less shared parameters which can not be well optimized without the proposed contrastive objective. Besides, after \our~ is applied, the performance on both tasks are comparable with either parameter-sharing or expert-sharing backbones. In this way, we can save the resource cost on designing and training complex gating mechanism.

\begin{table}[]
\caption{Performance comparison w.r.t. different backbones}
\begin{tabular}{ccccc}
\toprule
\multirow{2}{*}{Model} & \multicolumn{2}{c}{Click} & \multicolumn{2}{c}{Conversion} \\
                       & AUC         & GAUC        & AUC            & GAUC          \\ \midrule
Shared-Bottom~\cite{caruana1997multitask}          & 83.622      & 78.219      & 87.354         & 83.027        \\
\textbf{+\our}                 & \textbf{83.642}      & \textbf{78.229}      & \textbf{87.557}         & \textbf{83.265}        \\ \midrule
MMoE~\cite{Mmoe}                   & 83.642      & 78.256      & 87.393         & 83.006        \\
\textbf{+\our}                 & \textbf{83.667}      & \textbf{78.286}      & \textbf{87.569}         & \textbf{83.285}        \\ \midrule
PLE~\cite{PLE}                    & \textbf{83.660}      & \textbf{78.265}      & 87.188         & 82.798        \\
\textbf{+\our}                 & 83.652      & 78.255      & \textbf{87.515}         & \textbf{83.329}        \\ \bottomrule
\end{tabular}

\label{tab:exp-general}
\end{table}

\subsubsection{Performance on More Tasks.}
\begin{figure}[t]
    \begin{minipage}[t]{1\linewidth}
		\centering
		\includegraphics[width=1\textwidth]{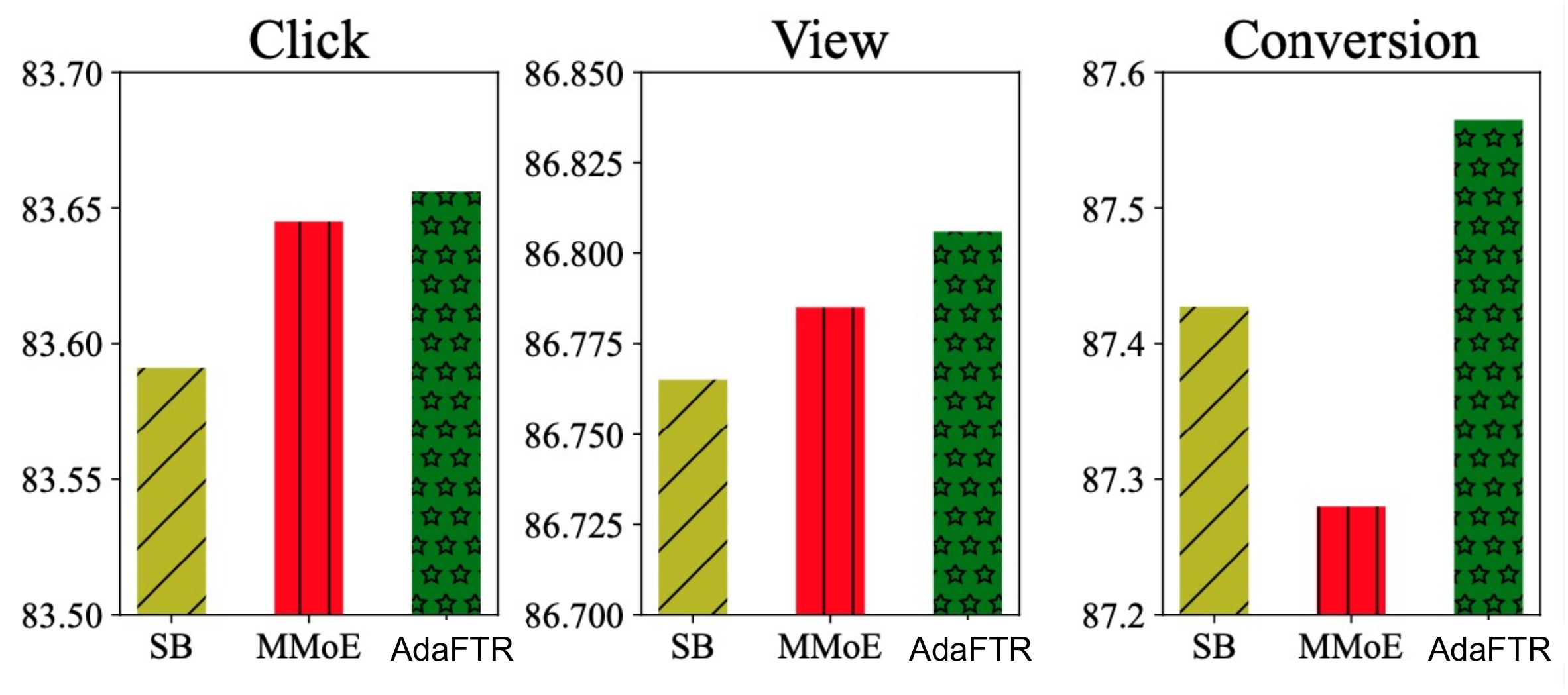}
		\subcaption{AUC} \label{fig:exp-three-tasks-auc}
	\end{minipage}
	\begin{minipage}[t]{1\linewidth}
		\centering
		\includegraphics[width=1\textwidth]{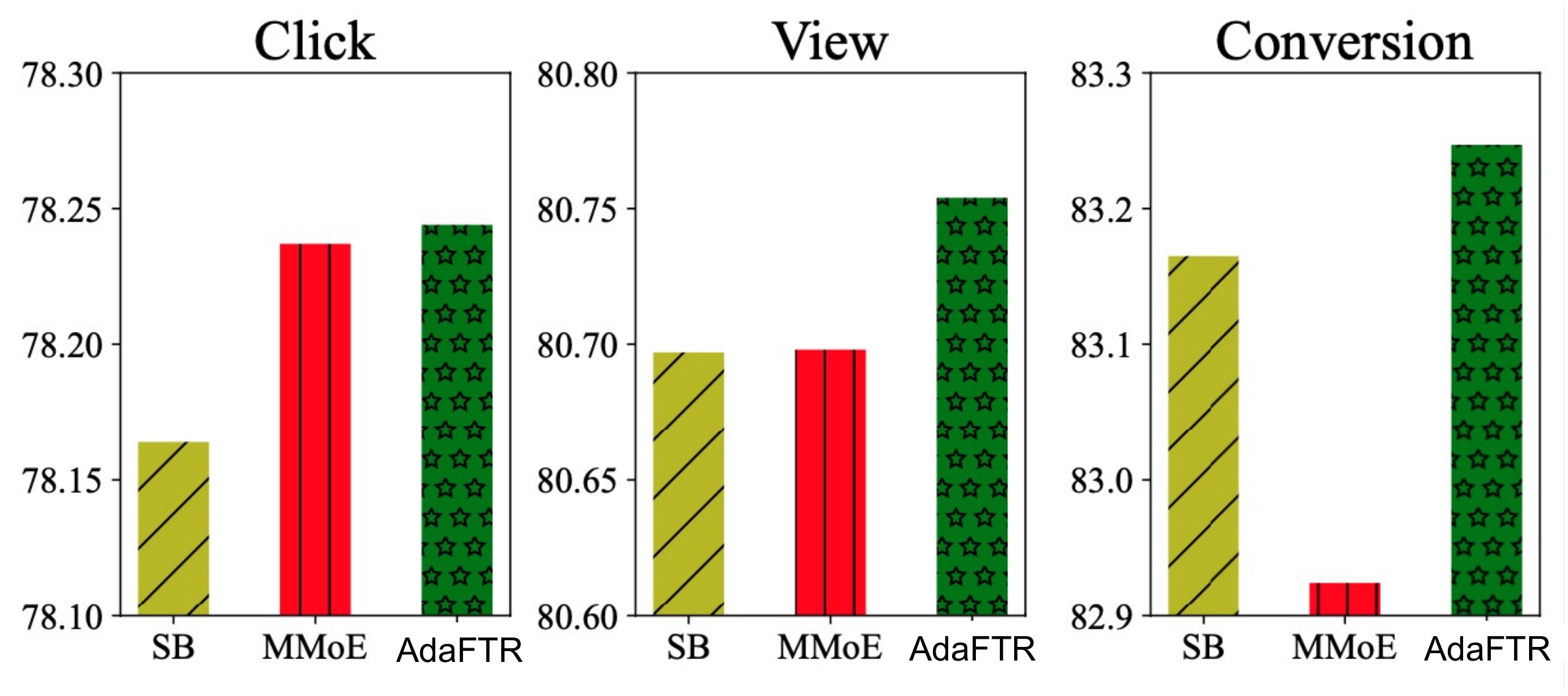}
		\subcaption{GAUC} \label{fig:exp-three-tasks-gauc}
	\end{minipage}
    \caption{Performance comparison of \our, Shared-Bottom (SB for short) and MMoE on three tasks~(\eg estimation on Click, View and Conversion). The results of AUC and GAUC are plotted at upper and lower, respectively.}
    \label{fig:exp-three-tasks}
\end{figure}

To verify the consistency of improvement on more than two tasks, we jointly consider three tasks of \emph{Click}, \emph{View} and \emph{Conversion} on Ecomm-Ads dataset where we regard the purchase of product as \emph{Conversion} and the browse of detail page as \emph{View}. Based on our experience and experimental results, the contrastive learning objectives are only conducted between Click and other tasks. The experimental results on three tasks are shown in Figure~\ref{fig:exp-three-tasks}. In the figure, we can find that the performance of \our~is consistently better than other baselines on all the tasks. Compared to the task of Conversion, the task of View has more instances with positive label~(\ie 1), which may bring more useful knowledge for assisting the learning of CTR Prediction. The overall performance on three tasks are much better compared with modeling only CTR and CVR tasks which shows the potential of introducing more tasks for modeling. Therefore, we can conclude that \our~can consistently improve the performance when more user behaviors could be collected and used for modeling.

\subsection{Online Deployment and A/B Test}

\begin{table}[]
\centering
\caption{Relative improvements in online A/B test. The lower the CPC, the better the performance.}
\begin{tabular}{@{}cccc@{}}
\toprule
Method & CTR~$\uparrow$ & CVR~$\uparrow$ & CPC~$\downarrow$ \\ \midrule
\textbf{\our} & \textbf{+2.37\%}  & \textbf{+4.82\%}    & \textbf{-5.78\%}      \\ \bottomrule
\end{tabular}


\label{tab:exp-online}
\end{table}

To evaluate the performance of \our, we deploy it in a large-scale financial advertising system at Alibaba with millions of user. Our algorithm aims to maximize the click probability when the poster is presented to users and the probability of purchase from both our advertisement and other links.  When the advertising requests come, the users portrait, candidate poster and other context feature would be sent to the model. And the predicted CTR and CVR will be utilized to guide the target price of the impression in the following function~\cite{he2021unified}:
\begin{equation}
    p_b = w \times \hat y_{ctr} \times \hat y_{cvr} \times p_c
\end{equation}
where $p_b$, $p_c$ and $w$ indicate the final price for advertising bidding, the statistical expected price and the weight for real-time calibration, respectively. And $\hat y_{ctr}, \hat y_{cvr}$ are the output of task-specific network in Equation~\ref{eq:formal} If our method can estimate CTR and CVR better, more cost will be paid for those high-value users. And the overall commercial profit can be improved.

The A/B test is conducted for 7 days with 40\% random traffic. We inherit the parameters in task-specific network from a well-trained DNN model and the network for task relatedness is optimized from scratch with the proposed \our method. And the overall model is updated daily. Table~\ref{tab:exp-online} reports the relative improvements on CTR, CVR and cost-per-conversion~(CPC) compared to the current online model. With insignificant cost on inference time, the superiority of \our~is consistent over all metrics, validating the effectiveness for industrial deployment. Moreover, the improvement on CVR is more significant which discloses the ability of \our~in promoting the performance of difficult task.
\section{Related work}
In this section, we briefly review related work in three aspects: CTR/CVR estimation, multi-task learning and contrastive learning.

\subsection{CTR/CVR Estimation}
In online recommender system and advertising systems, click-through rate~(CTR) and post-click conversion rate~(CVR) are core indicators of service~\cite{wang2020survey}.  From the traditional linear regression~\cite{richardson2007predicting} to latest neural-based models~\cite{zhang2021deep},  plenty of architectures were designed to either make in-depth feature interactions~\cite{guo2017deepfm,li2019fi} or incorporate representative data patterns~\cite{zhou2019deep,qin2020user}. Most prior studies leverage complex networks to improve the performance on a single task. For example, FiGNN~\cite{li2019fi} introduces graph neural networks to model feature interaction between different fields. Furthermore, AutoFIS~\cite{liu2020autofis} introduce learnable selection gates to make the feature interaction automatic and reduce redundant computation. 

Despite their effectiveness, those models may be sub-optimal since they omit the potential knowledge from other tasks. For instance, it has been proved that optimizing the both CTR and CVR estimation jointly can boost overall performance~\cite{AutoHERI}. 
Following this direction, hierarchical micro and macro behaviors are introduced to address the sample selection bias and data sparsity issues in CVR modeling~\cite{wen2021hierarchically}. And delicate parameter-sharing structures are designed in either expert level~\cite{zou2022automatic} or neuron level~\cite{xiao2020lt4rec} to exchange the information between CTR and CVR estimation network.

\subsection{Multi-Task Learning} For several related tasks, multi-task learning aims to jointly train a unified model with shared parameters or knowledge~\cite{collobert2008unified}. It has attracted more and more attention in the field of recommendation and advertising systems~\cite{yang2022cross,AutoHERI,pan2019predicting}. Some of these researches build the relations among tasks with business expertise~\cite{ma2018entire}, knowledge distillation~\cite{yang2022cross} or automatic architecture search~\cite{AutoHERI}. In another line of research,  well-designed architectures were designed to conduct knowledge sharing further~\cite{Mmoe,PLE,hazimeh2021dselect}. For example, MMoE~\cite{Mmoe} originally took multiple expert networks with learnable gates as the shared architecture. And PLE~\cite{PLE} further separated the gates into shared and specific ones.  For modeling multi-step conversion process of audiences, AITM~\cite{xi2021modeling} is proposed with adaptive information transfer module to learn. 

Different from these works, in our method, contrastive learning is conducted to model the inter-task relatedness. 
Instead of designing complex architectures, we just employ simple contrastive objectives to transfer the common knowledge and further update the learning with adaptive temperature. 
In addition, our paradigm is model-agnostic which can be combined with most of existing MTL architectures.

\subsection{Contrastive Learning} In light of the success of contrastive learning~\cite{he2020momentum}, researchers have devoted themselves to exploring its enormous potential in natural language processing~\cite{gao2021simcse}, graph mining~\cite{zhu2020deep} and recommendation~\cite{lin2022improving}. In most of the prior works, contrastive objective serve as auxiliary loss to convert traditional supervised learning into self-supervised multi-task learning~\cite{lin2020ms2l,lin2022improving}. Apart from the large-scale applications of contrastive learning, there are also several insightful works that analyse the underlying principles~\cite{wang2021understanding,robinson2021can}. Among them, an impressive summary is that alignment and uniformity are two vital characteristics that captured by contrastive learning~\cite{wang2020understanding}. 

To improve the task of CTR prediction, several insightful researches try to combine contrastive loss into the model training where feature embeddings are enhanced for sequential-based CTR tasks~\cite{guo2022miss}, cold-start scenarios~\cite{pan2021click} and general CTR prediction framework~\cite{wang2022cl4ctr}. 
In this work, instead of feature representations, we try to apply contrastive learning to achieve indirect knowledge exchange on the task representations for joint CTR-CVR  estimation and further upgrade it with instance-specific temperature for fine-grained relatedness~\cite{zhang2021temperature}.
\section{Conclusion}

This paper proposed  an \textbf{\underline{Ada}}ptive \textbf{\underline{F}}ine-grained \textbf{\underline{T}}ask \textbf{\underline{R}}elatedness modeling approach, \textbf{\our},  for joint CTR-CVR estimation.  Inspired by the MTL techniques, we design a hybrid architecture that combine both parameter sharing and representation alignment. Our backbone was developed based on the expert-sharing MTL architecture, and introduced a novel  adaptive inter-task representation alignment method based on contrastive learning.
We construct effective contrasting with  inter-task representations of a target instance and the representations of a random instance. 
The major novelty lies in the adaptive temperature in the InfoNCE loss of contrastive learning. To automatically set the temperature, we constructed a relatedness prediction network, so that it can predict the contrast strength for contrastive pairs. 
 In this way, our approach can better align the representation space of related tasks by using 
 adaptive contrast strength.  
Both \emph{offline evaluation} with public e-commerce datasets and \emph{online test} in a real advertising system at Alibaba have demonstrated the effectiveness of our approach. 

In this work, we examine the MTL setting with two classic prediction tasks, \ie CTR and CVR estimation. As future work, we will examine the feasibility of our approach in  other tasks, since the proposed approach is essentially task- and model-agnostic.   We will also explore the use of the adaptive contrast strength in general contrastive learning tasks.  

\ignore{a novel contrastive learning framework for multi-task learning that aims to achieve implicit knowledge integration between tasks.
We propose that, as opposed to designing hard or soft sharing architecture, the representations in the top task-specific networks can be contrasted.
Furthermore, we enhance the basic InfoNCE objective with customized temperatures to capture fine-grained relevance between tasks.
It is learned for each instance to regulate the degree of contrastive learning. 
The effectiveness and applicability of the proposed approach are demonstrated by extensive offline experiments on large-scale datasets and online A/B tests on industrial advertising systems.
}


\bibliographystyle{ACM-Reference-Format}

\end{document}